\documentclass[11pt]{article}
\usepackage{graphicx}
\usepackage{amsfonts}
\usepackage{amsmath}
\usepackage{amssymb}
\usepackage{url}
\usepackage{fancyhdr}
\usepackage{indentfirst}
\usepackage{enumerate}
\usepackage{amsthm}
\usepackage{color}
\usepackage{natbib}
\usepackage[colorlinks=true,citecolor=black]{hyperref}
\usepackage{subfigure}
\usepackage{threeparttable}
\usepackage{float}

\newcommand{\be}{\begin{eqnarray}}
\newcommand{\ee}{\end{eqnarray}}
\newcommand{\by}{\begin{eqnarray*}}
\newcommand{\ey}{\end{eqnarray*}}
\newcommand{\bn}{\begin{enumerate}}
\newcommand{\en}{\end{enumerate}}
\newcommand{\bi}{\begin{itemize}}
\newcommand{\ei}{\end{itemize}}

\renewcommand{\geq}{\geqslant}
\renewcommand{\leq}{\leqslant}
\renewcommand{\epsilon}{\varepsilon}
\theoremstyle{plain}
\newtheorem{theorem}{Theorem}

\newtheorem{proposition}[theorem]{Proposition}
\theoremstyle{definition}
\newtheorem{definition}{Definition}[section]

\theoremstyle{definition}

\numberwithin{equation}{section} \numberwithin{theorem}{section}
\renewcommand{\cite}{\citet}

\newcommand{\ignore}[1]{}

\begin{document}

\title{Wasserstein-Kelly Portfolios: A Robust Data-Driven Solution to Optimize Portfolio Growth}

\author{ Jonathan Yu-Meng Li \\
\\
Telfer School of Management\\
University of Ottawa \\
Ottawa, Ontario, Canada K1N 6N5\\
\\
}
\maketitle

We introduce a robust variant of the Kelly portfolio optimization model, called the Wasserstein-Kelly portfolio optimization. Our model, taking a Wasserstein distributionally robust optimization (DRO) formulation, addresses the fundamental issue of estimation error in Kelly portfolio optimization by defining a ``ball" of distributions close to the empirical return distribution using the Wasserstein metric and seeking a robust log-optimal portfolio against the worst-case distribution from the Wasserstein ball. Enhancing the Kelly portfolio using Wasserstein DRO is a natural step to take, given many successful applications of the latter in areas such as machine learning for generating robust data-driven solutions. However, naive application of Wasserstein DRO to the growth-optimal portfolio problem can lead to several issues, which we resolve through careful modelling. Our proposed model is both practically motivated and efficiently solvable as a convex program. Using empirical financial data, our numerical study demonstrates that the Wasserstein-Kelly portfolio can outperform the Kelly portfolio in out-of-sample testing across multiple performance metrics and exhibits greater stability.

\footnotetext{The author would like to thank Bob Barmish for bringing to the author's attention the question of how Wasserstein DRO may be applied to the Kelly portfolio problem, which inspired the research presented in this paper.}

\section{Introduction}
The Kelly strategy, developed by John Kelly in 1956 (\cite{K56}), has been influential in portfolio management. Originally designed to maximize cumulative wealth in repeated betting games, it has later been extended to determine the optimal allocation of an investor's capital to each investment at every time point, with the goal of maximizing the portfolio's expected growth rate in the long run (\cite{L59}). Various studies (\cite{K56}, \cite{B61}, \cite{AC88}) have demonstrated that a portfolio invested according to the Kelly strategy can outperform any other portfolio in the long run with certainty. Remarkably, this strategy requires only solving a single-period portfolio optimization problem to determine the optimal allocation of capital for the entire investment horizon, making it a myopic strategy. The portfolio optimization problem maximizes the expected logarithmic growth of the portfolio over a single time period, and it can be solved numerically through convex optimization techniques (\cite{BL04}).

The Kelly strategy assumes that investors have complete knowledge of the joint asset return distribution. However, in reality, the return distribution is not observable. One way to implement the Kelly strategy is to assume the functional form of the asset return distribution and calibrate it using historical data. However, this approach suffers from misspecification bias. Several studies (\cite{M89}, \cite{BG91}, and \cite{CZ93}) have shown that portfolios optimized based on a misspecified return distribution can perform poorly in out-of-sample testing. An alternative approach, which is fully data-driven, is to apply directly the empirical return distribution constructed from historical returns. This approach avoids the bias; however it raises the issue of over-fitting to historical samples, which can still lead to poor out-of-sample performance. Recent studies (\cite{RKW16}, \cite{SB18}, \cite{H22}) have focused on studying the Kelly strategy from a distributionally robust optimization (DRO) perspective, which offers a more robust solution to handle the uncertainty of asset return distribution.

The central idea behind DRO is to account for uncertainty in the data-generating distribution by constructing an ambiguity set of distributions that can include the data-generating distribution with high confidence. A robust solution is obtained by optimizing against the worst-case distribution in the ambiguity set. \cite{RKW16} studied the problem of 
robust growth-optimal portfolio by applying an ambiguity set constructed from the first two moments of asset returns. Their solution is robust against any distribution with the same first two moments as those estimated from historical returns but may be overly conservative as it does not incorporate any other distributional information beyond the first two moments. Recent studies by \cite{SB18} and \cite{H22} have taken a data-driven approach to DRO by defining an ambiguity set that shares the same support as the empirical return distribution but with uncertain mass probabilities. \cite{SB18} used two probability metrics, f-divergence and Wasserstein distance, to measure the distance between two discrete distributions with the same support. They defined the distributions in an ambiguity set as any distribution that is within a certain distance from the empirical distribution according to the probability metrics. This approach is data-driven in that one can reduce the distance parameter as more samples become available and the empirical distribution becomes increasingly accurate, causing the ambiguity set to converge asymptotically to the data-generating distribution.

The approach taken by \cite{SB18} has a significant limitation that it only applies to distributions with finite support. As asset return distributions are continuous by nature, the ambiguity sets used in their approach would not be able to contain the true asset return distribution. This could be a concern for investors who are worried about extreme events that may cause assets to deviate significantly from their historical performances. However, recent developments in DRO have addressed this limitation by using the Wasserstein metric to define an ambiguity set, known as the Wasserstein ball, applicable to both discrete and continuous distributions. Specifically, \cite{EK18} showed that distributionally robust optimization problems with Wasserstein balls, now known as Wasserstein DRO, can be solved efficiently as convex programs for many optimization problems and enjoy finite-sample performance guarantees.  Further research has explored other advantages of Wasserstein DRO, such as generalization bounds and regularization effects, as demonstrated in recent studies by \cite{G22} and \cite{WLM22}.

Our paper demonstrates how Wasserstein DRO can be applied to solve the Kelly portfolio optimization problem, resulting in a Kelly portfolio that is driven by return data but protected against overfitting. We note that a naive application of Wasserstein DRO to the Kelly portfolio problem, by defining an ambiguity set for the distribution of simple returns, as done in earlier works (\cite{RKW16}, \cite{SB18}, \cite{H22}), would lead to an inadequate model that provides no meaningful solutions. The key to successfully applying Wasserstein DRO to the Kelly portfolio problem, as shown in this paper, is to define a Wasserstein ball for the primary source of uncertainty - the distribution of log-returns. This choice is well-justified from the perspective of financial modelling but has been less explored in the application of DRO to portfolio optimization. Our approach leads to a novel Wasserstein DRO model, the Wasserstein-Kelly portfolio optimization model, which not only better captures the asymmetrical nature of asset return distribution but also admits reformulation as a finite-dimensional convex program. This reformulation can be efficiently solved using off-the-shelf solvers. We demonstrate in Section 3 how Wasserstein-Kelly portfolios diversify the Kelly portfolios and present the results of an extensive set of experiments using data from the S\&P500. Our results show that the Wasserstein-Kelly portfolio can outperform the Kelly portfolio across multiple performance metrics and exhibit greater stability.

\section{Wasserstein-Kelly Portfolio}
\subsection{Kelly Strategy}
In this section, we start by introducing the basic setup of multi-period investing and the growth-optimal portfolio problem, which is a key concept in the Kelly strategy. Given a financial market with $n$ assets, we denote the prices of these assets at some future time $t \in \{1,...T\}$ by a random vector $\vec{S}_{t}:=(S_{t,1},...,S_{t,n})$ and their respective simple returns by $\vec{R}_{t}:=(R_{t,1},...,R_{t,n})$, i.e. 
$$R_{t,i}=\frac{S_{t,i}-S_{t-1,i}}{S_{t-1,i}},\;t=1,...,T,\;i=1,...,n.$$
At each time point $t$, investors can adjust the distribution of their wealth among $n$ different assets through portfolio rebalancing. This involves determining the percentage of their portfolio to allocate to each asset. We denote this decision by a wealth allocation vector $w_{t}=(w_{t,1},...,w_{t,n})$, which is non-negative and summed up to 1. For convenience, we write $w_{t}\in{\cal W}$, where ${\cal W}:=\left\{ w \in \mathbb{R}^n \;|\;\vec{1}^{\top}w=1,\;w\geq0\right\}$ . 

A portfolio strategy refers to all the decisions $\{w_{t}\}_{t=1}^{T}$ that should be made over time, which can depend on the market information available at the time of each decision. Assuming that the initial capital of investors is one, we can calculate the final portfolio value at time $T$ for a strategy by 
$$V_{T}=\prod_{t=1}^{T}\left(1+\vec{R}_{t}^\top w_t \right).$$
The growth rate of the portfolio is defined as the natural logarithm of the geometric mean of the absolute returns, i.e. 
$$\gamma_{T}=\log\sqrt[T]{\prod_{t=1}^{T}\left(1+\vec{R}_{t}^\top w_{t} \right)}=\frac{1}{T}\sum_{t=1}^{T}\log\left(1+\vec{R}_{t}^\top w_{t}\right).$$

The objective of the growth-optimal portfolio problem is to identify a strategy that maximizes the portfolio's growth rate. One natural class of strategies that can be considered is the constant proportions strategies, where investors determine the optimal composition of the portfolio to maintain throughout the entire investment horizon. These strategies can be reduced to setting the portfolio composition to a fixed value $w \in {\cal W}$ that remains constant over the entire investment horizon, i.e. $w_t = w, \forall t$. Following the strong law of large numbers, the growth rate of the portfolio for any constant proportions strategy in the long run would satisfy 
\begin{equation}
\lim_{T\rightarrow\infty}\gamma_{T}=\mathbb{E}\left[\log\left(1+\vec{R}_{t}^\top w\right)\right], \label{ke}
\end{equation}
under the assumption that the simple returns $\vec{R}_{t}$ are white noise, i.e. mutually independent and identically distributed. 

By exploiting this fact, Kelly identified that the optimal constant proportion strategy that maximizes the asymptotic growth rate of the portfolio, under the white noise assumption, can be obtained simply by maximizing the right-hand-side of \eqref{ke}, i.e. 
\begin{equation}
\max_{w\in{\cal W}}\;\mathbb{E}\left[\log\left(1+\vec{R}^\top w \right)\right]. \label{kmodel}
\end{equation}
Here, $\vec{R}$ is a random vector sharing the same distribution as any $\vec{R}_t$. The Kelly strategy is essentially solving the above single-period Kelly portfolio optimization model, which is a convex optimization problem (\cite{BL04}).

\subsection{Wasserstein-Kelly Portfolio Optimization}
The Kelly strategy assumes that investors know the probability distribution of asset returns, denoted by $\vec{R} \sim F_R$, which is required to solve the problem $\eqref{kmodel}$. However, in practice, investors typically only have access to a finite set of sample returns denoted by $\hat{R}_1, \ldots, \hat{R}_N$. To apply the Kelly strategy, a simple approach is to use the empirical distribution 
$$\hat{F}_{R}:=\frac{1}{N}\sum_{j=1}^{N}\delta_{\hat{R}_{j}},$$
where $\delta$ is the delta function, to replace $F_R$ and instead solve an empirical version of $\eqref{kmodel}$, given by
\begin{equation}
\max_{w\in{\cal W}}\frac{1}{N}\sum_{j=1}^{N}\log\left(1+\hat{R}_{j}^\top w\right).\label{emodel}
\end{equation}
While this approach, also known as sample average approximation (SAA), is fully data-driven and does not require assumptions about the distributional form of $F_R$, its major limitation is well known -- it is sensitive to sampling errors and suffers from the issue of overfitting. 

We demonstrate in this section how the method of Wasserstein distributionally robust optimization (DRO) can be applied to overcome this limitation. As preliminaries, we recall first the definition of the type-$p$ Wasserstein metric and Wasserstein ball. 
\begin{definition}
Given any two distributions $F_{1}$ and $F_{2}$, the type-$p$ Wasserstein metric is defined by
$$
d_{p}(F_{1},F_{2}):=\inf_{\Pi}\;\left\{ \left(\mathbb{E}\left[||\xi_{1}-\xi_{2}||^{p}\right] \right)^{1/p} \middle |\begin{array}{c}
\Pi\;\text{is a joint distribution of }\text{\ensuremath{\xi_{1}}and \ensuremath{\xi_{2}}}\\
\text{with marginals }F_{1}\text{ and }F_{2},\text{respectively.}
\end{array}\right\},$$
for any $p \in [1,\infty)$.
\end{definition}
The type-$p$ Wasserstein metric can be applied to build a set of distributions, i.e. the Wasserstein ball, that include any distribution similar to the empirical distribution. Suppose that $\hat{\xi}_{1},...,\hat{\xi}_{N}$ represent $N$ random samples drawn from a general finite-dimensional random vector $\xi \sim F_{\xi}$ and let $\hat{F}:=\frac{1}{N}\sum_{j=1}^{N}\delta_{\hat{\xi}_{j}}$ denote the empirical distribution. The type-$p$ Wasserstein ball is defined by 
$${\cal B}_{p}(\varepsilon):=\left\{ F\; \middle |\;d_{p}(F,\hat{F})\leq\varepsilon\right\}, $$
where $\varepsilon >0$ is the radius of the ball, i.e. the maximum distance between any distribution in the ball and the empirical distribution. The application of the type-$p$ Wasserstein ball to capture the uncertainty of the data-generating distribution $F_{\xi}$ is supported by the following statistical guarantees. 

\begin{theorem} (\cite{FG15}) Let $\hat{\xi}_{1},...,\hat{\xi}_{N}$ be $N$ samples drawn independently from a distribution $F_{\xi}$ and $\mathbb{P}^N$ denote the distribution that governs the distribution of the independent samples. Assuming that the distribution $\xi \sim F_{\xi}$ is a light-tailed distribution such that $\mathbb{E}[\exp(||\xi||^\alpha)] < \infty$ for some $\alpha > p$, $p \geq 1$, then for any $\varepsilon \geq 0$, 
$$\mathbb{P}^N \left[  d_{p}(F_{\xi},\hat{F}) > \varepsilon \right] \leq h_p(\varepsilon)$$ 
for some $h_p : \mathbb{R}_{+} \rightarrow \mathbb{R}_{+}$ that decreases to zero at an exponential rate in $\varepsilon$.
\end{theorem}

Wasserstein DRO refers to an optimization formulation that seeks to generate a robust solution by optimizing against the worst-case distribution from the type-$p$ Wasserstein ball. One natural, yet naive, attempt to applying Wasserstein DRO to the Kelly portfolio optimization problem is to build the type-$p$  Wasserstein ball for the distribution of simple returns $F_R$, i.e.
$${\cal B}_{p}(\varepsilon):=\left\{ F\; \middle |\;d_{p}(F,\hat{F}_R)\leq\varepsilon\right\}. $$
This leads to the following Wasserstein DRO formulation of the Kelly portfolio optimization problem:
\begin{equation}
\max_{w\in{\cal W}}\inf_{F_R \in {\cal B}_{p}(\varepsilon)} \mathbb{E}_{F_R}\left[\log(1+ \vec{R}^\top w)\right]. \label{eq:model1}
\end{equation}
Upon closer examination of the above Wasserstein DRO problem, several issues with the model become apparent. Firstly, the support of a distribution $F_R$ from the Wasserstein ball, i.e. $F_R \in {\cal B}_{p}$, may be unbounded. This means that the absolute return $1+ \vec{R}^\top w$ may be negative, rendering it incompatible with the domain of the logarithmic function. Secondly, even if a constraint is imposed to limit the support of distributions $F_R$ to be non-negative, the inner minimization problem, which is the worst-case expectation problem, is generally unbounded, even when the radius $\varepsilon$ is small. Thirdly, and perhaps most importantly, it is questionable whether the uncertainty of simple returns is the true primary source of uncertainty for the random behaviour of asset prices. 

\subsubsection*{Wasserstein-Kelly portfolio optimization model}
It turns out that all the aforementioned issues can be resolved by focusing instead on the distribution of the log-returns, i.e. $$r_{t}=\ln\frac{S_{t}}{S_{t-1}}.$$
This is actually aligned with how the random behavior of stock prices is commonly modeled in the finance literature.  We denote the log-returns of $n$ assets by $\vec{r}_{t}:=(r_{t,1},...,r_{t,n})=\left(\ln\frac{S_{t,1}}{S_{t-1,1}},...,\ln\frac{S_{t,n}}{S_{t-1,n}}\right).$ Note that the absolute return of the portfolio can be written as 
$$1+\vec{R}_{t}^\top w=\exp\left(\vec{r}_{t}\right)^\top w,$$
where $\exp\left(\vec{r}_{t}\right) := \left(\exp\left(r_{t,1}\right),...,\exp\left(r_{t,n}\right)\right).$

Now, let $\hat{r}_{1},...,\hat{r}_{N}$ represent $N$ samples of log-returns and $\hat{F}_{r}:=\frac{1}{N}\sum_{j=1}^{N}\delta_{\hat{r}_{j}}$ denote the empirical distribution of log-returns. We can build the type-$p$ Wasserstein ball for the distribution of log-returns, i.e. 
$${\cal B}_{p}(\varepsilon):=\left\{ F\; \middle |\;d_{p}(F,\hat{F}_r)\leq\varepsilon\right\}, $$
and consider the following Wasserstein DRO formulation of the Kelly portfolio optimization problem:
\begin{equation}
\max_{w\in{\cal W}}\inf_{F_r \in {\cal B}_{p}(\varepsilon)} \mathbb{E}_{F_r}\left[\log\left(\exp\left(\vec{r}\right)^\top w\right)\right]. \label{eq:model2}
\end{equation}

The above model, referred to as Wasserstein-Kelly portfolio optimization model, is a novel Wasserstein DRO formulation that has not appeared in the literature of Wasserstein DRO. Our choice of modelling the uncertainty of the log-return distribution, leading to the introduction of the exponential terms to the objective function in \eqref{eq:model2}, turns to be critical also from a computation perspective. It allows for converting the problem to a more tractable optimization problem. Specifically, it ``convexifies" the objective function with respect to the variable $\vec{r}$, enabling a further reformulation of the model into a finite-dimensional convex program, which can be solved via off-the-shelf solvers. We defer the proof of the following reformulation result to the appendix.

\begin{proposition} \label{main}
The Wasserstein-Kelly portfolio optimization model \eqref{eq:model2} can be solved via the following convex optimization program:
\begin{equation}
\max_{w \in {\cal W},v^{(j)}\geq0,\lambda\geq0}\;\frac{1}{N}\sum_{j=1}^{N}\left(\hat{r}_{j}^{\top}v^{(j)}-(p-1)p^{-(\frac{p}{p-1})}\lambda||\frac{v^{(j)}}{\lambda}||_{*}^{\frac{p}{p-1}}+\sum_{i=1}^{n}v_{i}^{(j)}\log\left(\frac{w_{i}}{v_{i}^{(j)}}\right)\right)-\lambda\varepsilon^{p},
\end{equation}
where $w\in\mathbb{R}^{n},\nu^{(j)}\in\mathbb{R}^{n},\lambda\in\mathbb{R}$
, and $||\cdot||_{*}$ denotes the dual norm of the norm $||\cdot||$. For $p=1$, the program is equivalent to 
\begin{equation} \nonumber
\max_{w \in {\cal W},v^{(j)}\geq0,\lambda\geq||v^{(j)}||_{*}}\;\frac{1}{N}\sum_{j=1}^{N}\left(\hat{r}_{j}^{\top}v^{(j)}+\sum_{i=1}^{n}v_{i}^{(j)}\log\left(\frac{w_{i}}{v_{i}^{(j)}}\right)\right)-\lambda\varepsilon.
\end{equation}
\end{proposition}

\section{Numerical Studies}
In this section, we evaluate the effectiveness of the Wasserstein-Kelly portfolio optimization model by comparing it to the Kelly portfolio using historical financial data from the stocks listed in the S\&P500 index between 2019/01/01 and 2023/02/20. We first demonstrate the impact of the radius-size parameter on portfolio diversification using a fixed set of stocks, and then assess the out-of-sample performance of the two portfolios across a large number of randomly selected stocks from the indices. Throughout our numerical studies, we use the Wasserstein-Kelly model based on the type-2 Wasserstein metric (i.e., $p$=2), which is a natural choice if one assumes the distribution of log-return has finite first and second moments. Both models are implemented using the python package CVXPY.

\subsection{Diversification Effects}
\begin{figure}[H]
  \centering
  \includegraphics[width=0.5\textwidth]{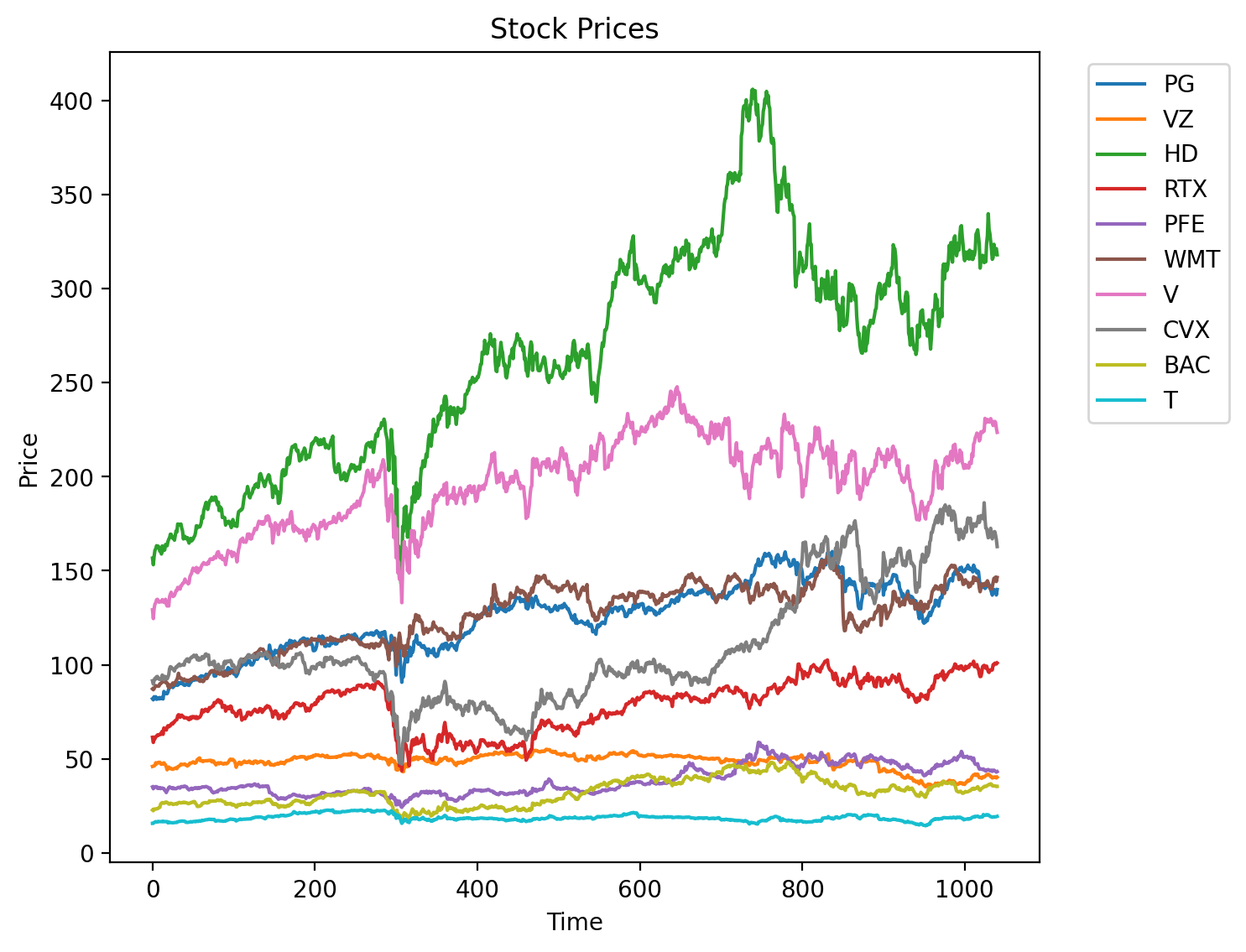}
  \caption{The stock prices of the ten stocks}
  \label{price}
\end{figure}

\begin{figure}[H]
  \centering
  \includegraphics[width=0.5\textwidth]{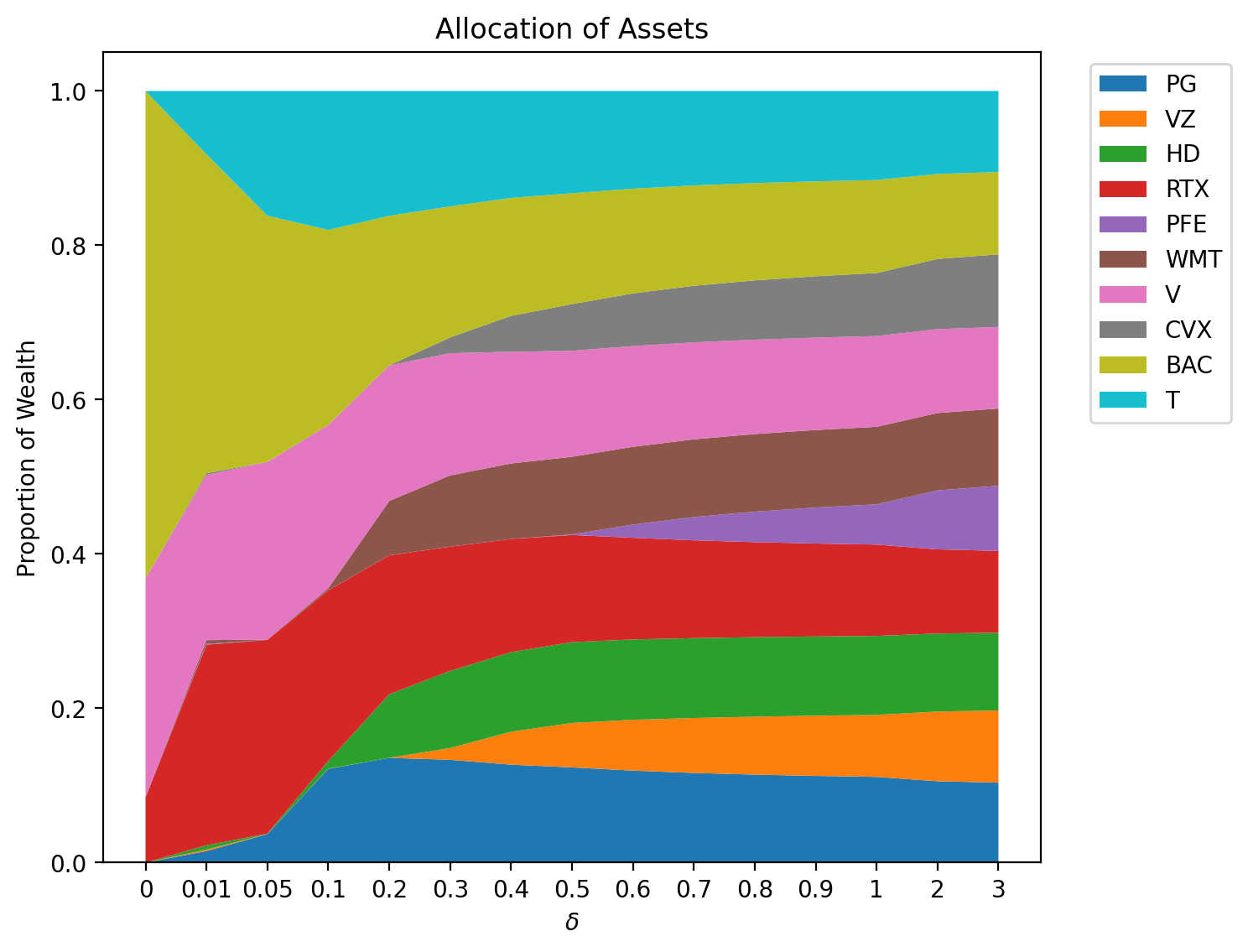}
  \caption{The impact of varying the proportion parameter $\delta$ on the composition of the Wasserstein-Kelly portfolio}
  \label{comp}
\end{figure}

We investigate the impact of varying the radius parameter, $\varepsilon$, on the composition of the Wasserstein-Kelly portfolio. We set $\varepsilon$ as a multiple of the average log-return, denoted by $\bar{R}$, across all assets, specifically $\varepsilon=\delta\cdot\bar{R}$, where $\delta$ is a proportion parameter that determines the size of $\varepsilon$ relative to $\bar{R}$. For example, choosing $\delta=0.1$ corresponds to allowing for a 10\% perturbation to the historical returns. We present the prices of ten stocks used in this section in Figure \ref{price}, while Figure \ref{comp} shows the changes in portfolio composition for a range of $\delta$ values. Increasing $\delta$ results in greater portfolio diversification, as illustrated in Figure \ref{comp}, and the portfolio converges to the $1/N$ portfolio in the limit. This showcases how the Wasserstein-Kelly portfolio achieves robustness through diversification and how it relates to the Kelly portfolio and $1/N$ portfolio, which represent the two extremes. Note that the Wasserstein-Kelly portfolios presented in Figure \ref{comp} are optimized using the first 252 days data in Figure \ref{price}.

\subsection{Out-of-Sample Performances}
In this section, we conduct experiments to compare the out-of-sample performances of the Kelly portfolio and the Wasserstein-Kelly portfolio. We randomly select 10 stocks from the S\&P500 and compute the Kelly portfolio and the Wasserstein-Kelly portfolio using historical data from the year of 2019 (252 trading days). We then evaluate their out-of-sample performance using data from 2020/01/01 to 2023/02/20. This process is repeated 1000 times with different randomly selected assets for portfolio optimization. We report results for the Kelly portfolio and four Wasserstein-Kelly portfolios with $\delta$ set to $0.1, 0.2, 0.3, 0.4$.

\begin{figure}[H]
  \centering
  \includegraphics[width=0.6\textwidth]{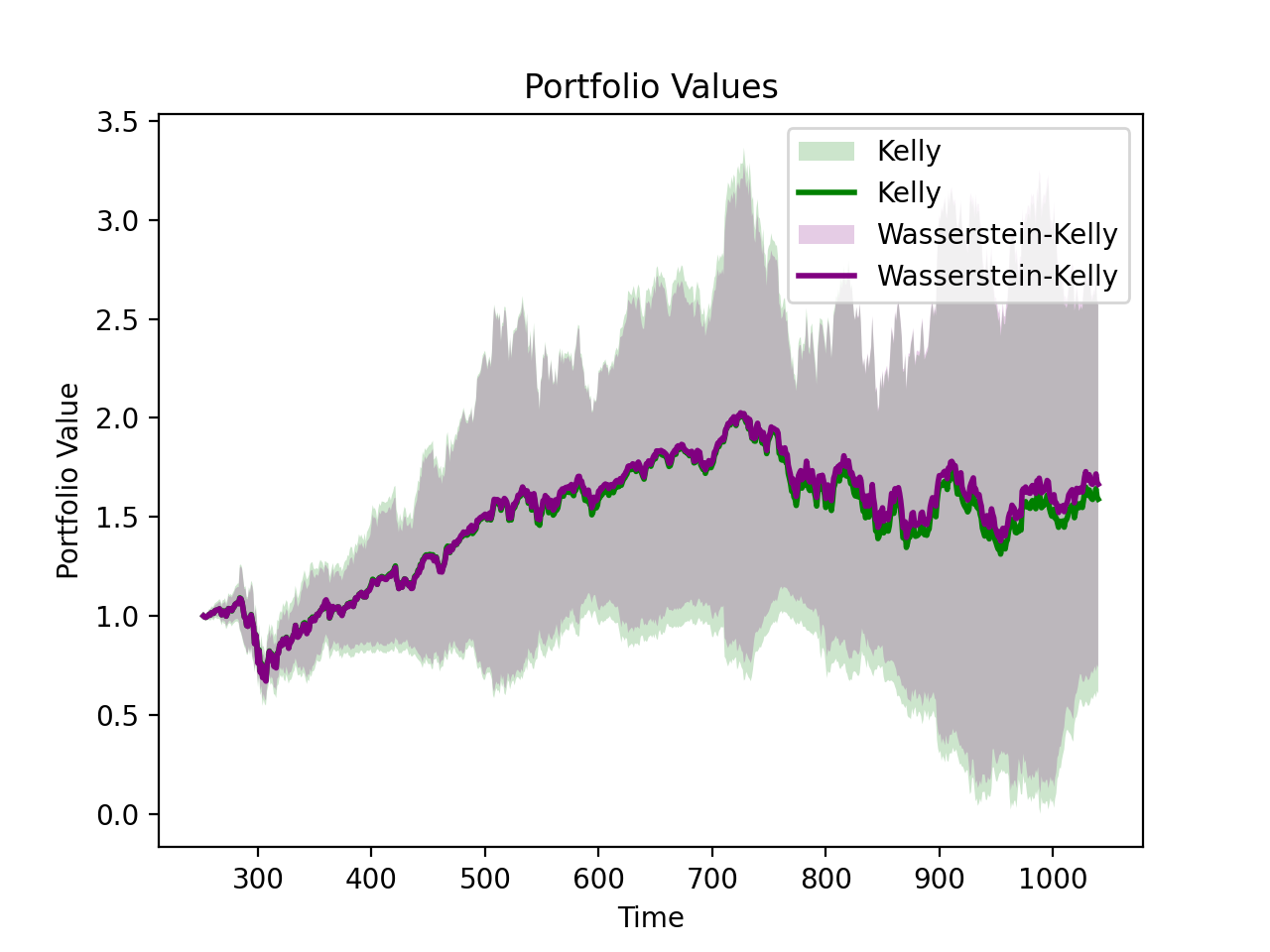}
  \caption{The evolution of portfolio values for the Kelly portfolio and the Wasserstein-Kelly portfolio with $\delta = 0.1$ over the period 2020/01/01-2023/02/20 (bold: average, shaded: standard deviation).}
  \label{pv0}
\end{figure}

\begin{figure}[H]
  \centering
  \includegraphics[width=0.6\textwidth]{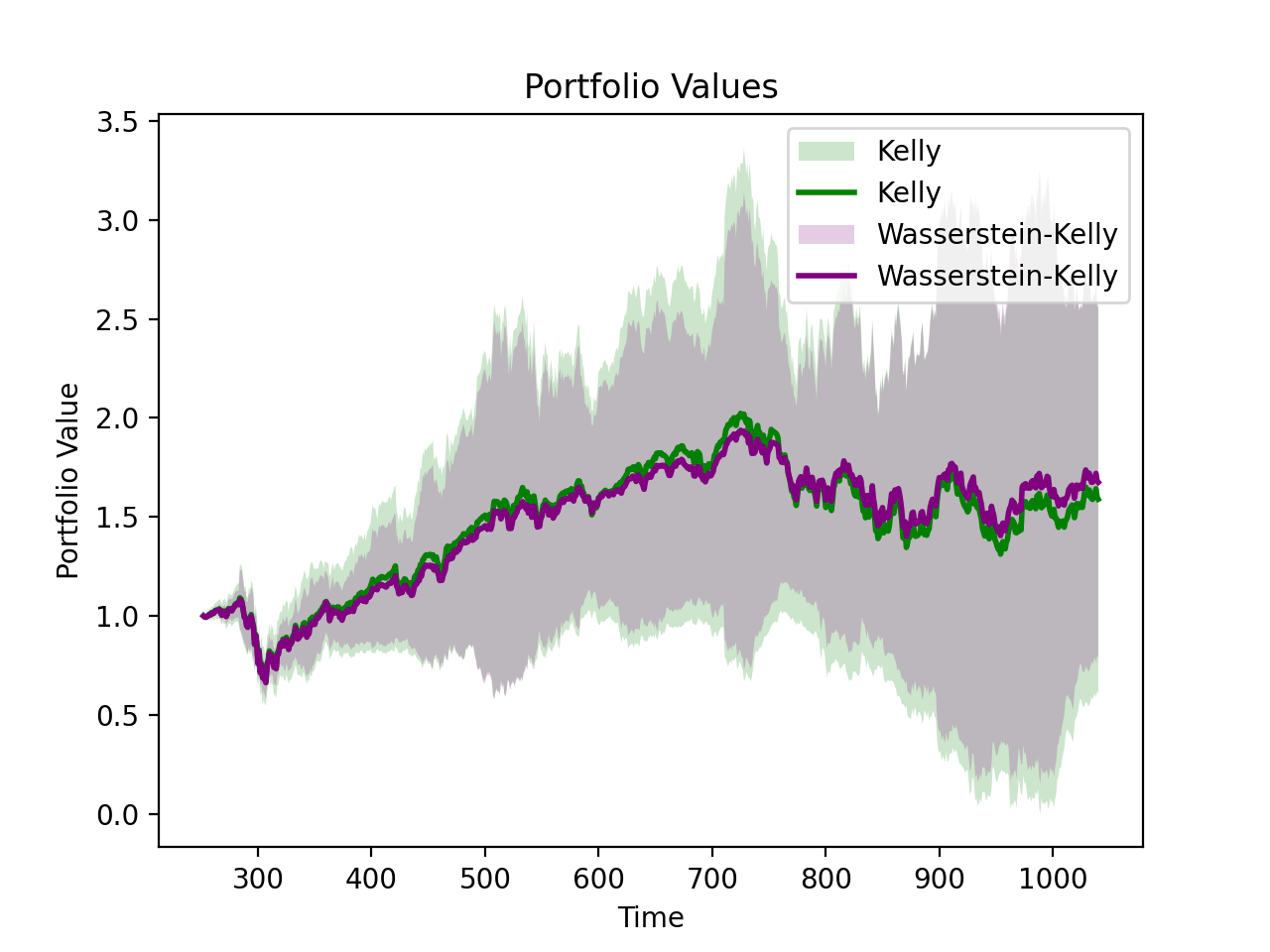}
  \caption{The evolution of portfolio values for the Kelly portfolio and the Wasserstein-Kelly portfolio with $\delta = 0.2$ over the period 2020/01/01-2023/02/20 (bold: average, shaded: standard deviation).}
  \label{pv1}
\end{figure}

\begin{figure}[H]
  \centering
  \includegraphics[width=0.6\textwidth]{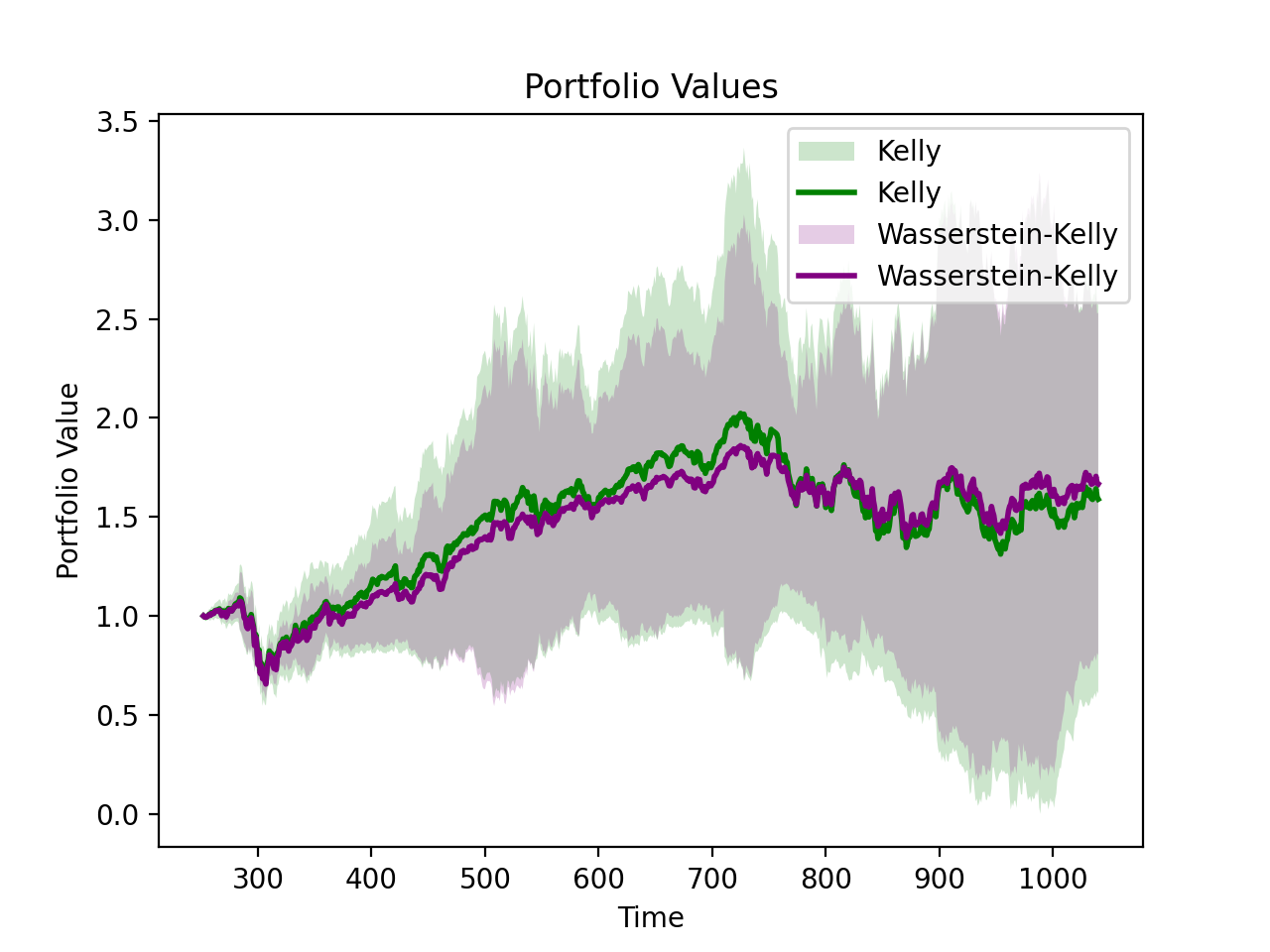}
  \caption{The evolution of portfolio values for the Kelly portfolio and the Wasserstein-Kelly portfolio with $\delta = 0.3$ over the period 2020/01/01-2023/02/20 (bold: average, shaded: standard deviation).}
  \label{pv2}
\end{figure}

\begin{figure}[H]
  \centering
  \includegraphics[width=0.6\textwidth]{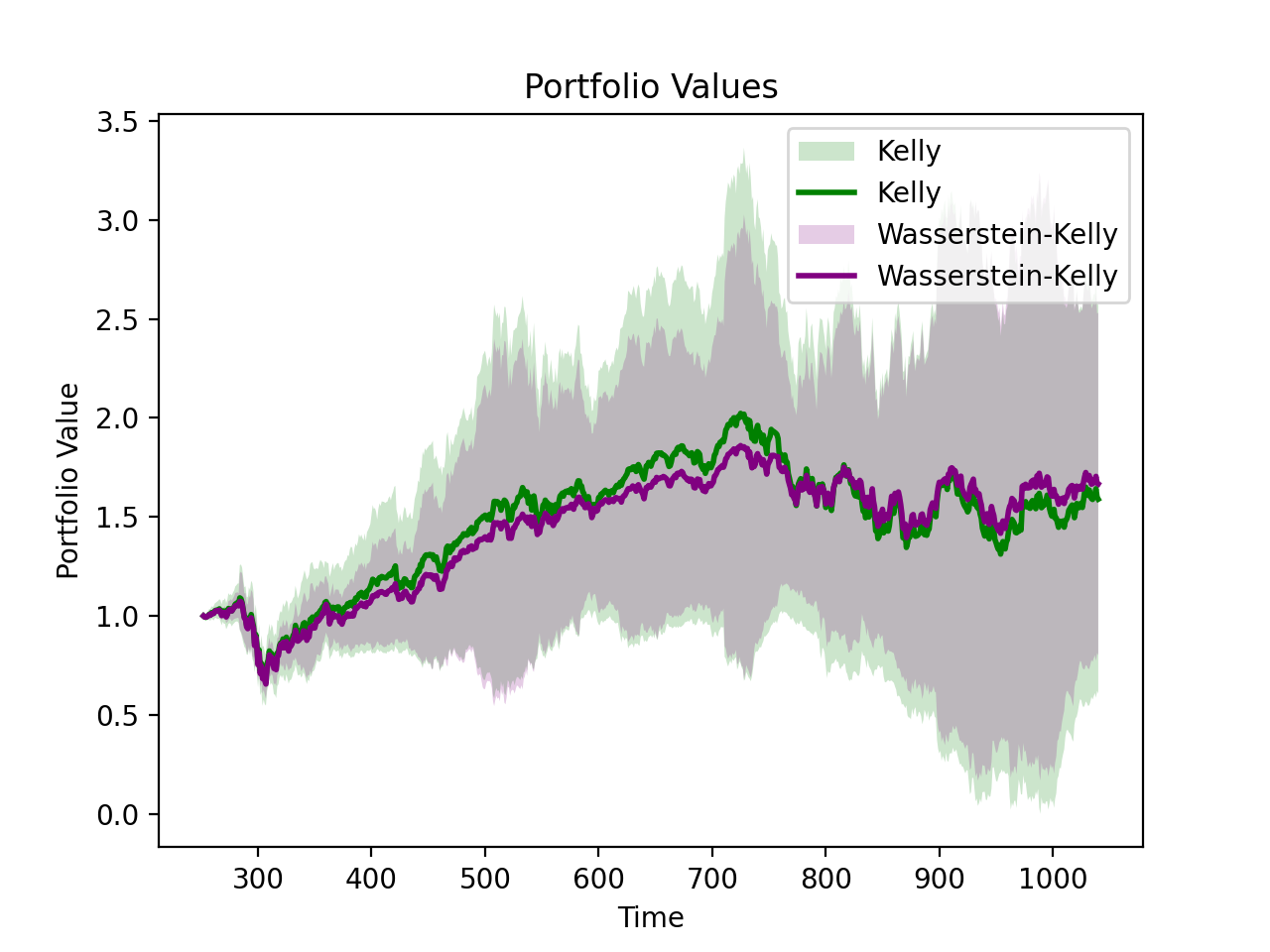}
  \caption{The evolution of portfolio values for the Kelly portfolio and the Wasserstein-Kelly portfolio with $\delta = 0.4$ over the period 2020/01/01-2023/02/20 (bold: average, shaded: standard deviation).}
  \label{pv3}
\end{figure}

\begin{figure}[H]
  \centering
  \includegraphics[width=1.0\textwidth]{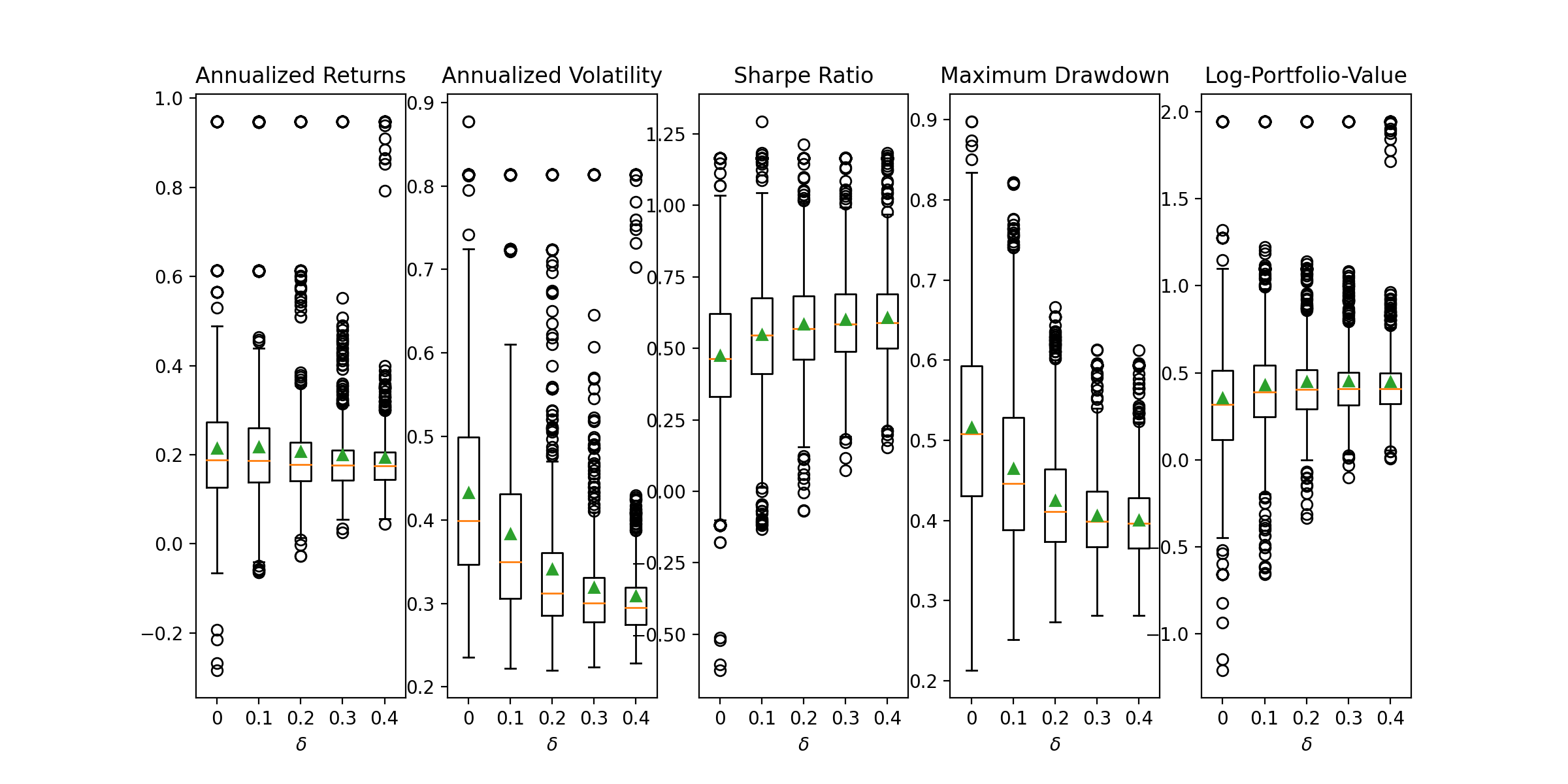}
  \caption{Boxplots of Kelly and Wasserstein-Kelly portfolios for various performance metrics: each subplot summarizes the performance of a portfolio based on 1000 randomly selected stocks. Performance metrics include annualized returns, annualized volatility, Sharpe ratio, maximum drawdown, log of portfolio value (at the end of investment horizon), and $\delta=0$ corresponds to the Kelly portfolio, while $\delta = 0.1, 0.2, 0.3, 0.4$ correspond to the Wasserstein-Kelly portfolios. The green triangle indicates the average performance.}
  \label{stats}
\end{figure}

We present visual representations of portfolio values over the out-of-sample period 2020/01/01-2023/02/20 for both Kelly and Wasserstein-Kelly portfolios in Figures \ref{pv0}-\ref{pv3}. Each figure compares the Kelly portfolio against the Wasserstein-Kelly portfolio with a different $\delta$ value. The bold lines represent the average of portfolio values over 1000 trajectories (each corresponding to a randomly selected set of 10 stocks used to build the portfolio), while the shaded regions depict the standard deviation of the portfolio values over these trajectories. Boxplots in Figure \ref{stats} provide detailed statistics, including annualized return, volatility, Sharpe ratio, maximum drawdown, and log of final portfolio value.

Our results indicate that the Wasserstein-Kelly portfolios exhibit a more stable evolution of portfolio values than the Kelly portfolio. The standard deviation (indicated by the shaded region) of the former is always less than that of the latter, with larger $\delta$ values leading to more stable performance. Additionally, the Wasserstein-Kelly portfolio outperforms the Kelly portfolio across all important performance metrics, including annualized returns, volatility, Sharpe ratio, maximum drawdown, and log of final portfolio value. Notably, the boxplots of the Wasserstein-Kelly portfolio ``shrink" as the value of $\delta$ increases, demonstrating how increasing robustness can reduce performance uncertainty.


\section{Conclusion}
In this paper, we introduce the Wasserstein-Kelly portfolio optimization model, which leverages a novel Wasserstein DRO formulation to enhance the Kelly portfolio. Our study underscores the importance of careful modelling in applying the Wasserstein DRO to the Kelly portfolio optimization problem. By utilizing the Wasserstein ball to model the uncertainty of log-return distribution, we arrive at a model that is both practically well-motivated and computationally efficient. Our numerical study demonstrates that the Wasserstein-Kelly portfolio is a promising alternative to the Kelly portfolio, particularly for investors concerned about the volatility of the Kelly portfolio and the noise in historical data. Overall, our findings highlight the potential of the Wasserstein-Kelly portfolio as a valuable tool in portfolio optimization.

\section{Appendix}
\subsection*{Proof of Proposition \ref{main}}
\begin{proof}
The optimization problem \eqref{eq:model2} can be written as 
\begin{align*}
\max_{w\in{\cal W}}\inf_{\Pi,F_{r}} & \;\int\log\left(\sum_{i=1}^{n}e^{r_{i}}w_{i}\right)dF_{r}\\
{\rm s.t.} & \;\int||r-r'||^{p}d\Pi\leq\varepsilon^{p}\\
 & \;\begin{array}{c}
\Pi\;\text{is a joint distribution of }\text{\ensuremath{r}\;and \ensuremath{r'}}\\
\text{with marginals }F_{r}\text{ and }\hat{F}_{r},\text{respectively.}
\end{array}
\end{align*}
By the strong duality result of Wasserstein DRO (see e.g., \cite{GK16})), the inner minimization problem can be equivalently formulated as the following maximization problem
\begin{align}
 & \sup_{\lambda\geq0}\;\int\left[\inf_{r\in\mathbb{R}^{n}}\left\{ \log\left(\sum_{i=1}^{n}e^{r_{i}}w_{i}\right)+\lambda||r-r'||^{p}\right\} \right]dF_{r'}-\lambda\varepsilon^{p}\nonumber \\
= & \sup_{\lambda\geq0}\;\frac{1}{N}\sum_{j=1}^{N}\left[\inf_{r\in\mathbb{R}^{n}}\left\{ \log\left(\sum_{i=1}^{n}e^{r_{i}}w_{i}\right)+\lambda||r-\hat{r}_{j}||^{p}\right\} \right]-\lambda\varepsilon^{p}.\label{eq:dual}
\end{align}
We focus first on the reformulation of the inner minimization problem,
i.e. 

\begin{equation}
\inf_{r\in\mathbb{R}^{n}}\left\{ \log\left(\sum_{i=1}^{n}e^{r_{i}}w_{i}\right)+\lambda||r-\hat{r}_{j}||^{p}\right\} .\label{eq:mini}
\end{equation}
Note that given any $w\in{\cal W}$ and $\lambda\geq0$, both the
term
\begin{equation}
f(r):=\log\left(\sum_{i=1}^{n}e^{r_{i}}w_{i}\right)
\end{equation}
and the term
\begin{equation}
g(r):=\lambda||r-\hat{r}_{j}||^{p}
\end{equation}
are convex functions. We can thus apply the Fenchel duality theorem
to the problem (\ref{eq:mini}), giving us
\begin{align}
 & \inf_{r\in\mathbb{R}^{n}}f(r)+g(r)\nonumber \\
= & \max_{v}-f^{*}(v)-g^{*}(-v)\label{eq:conjugate}
\end{align}
where $f^{*}$ and $g^{*}$ denote the convex conjugates of $f$ and
$g$ respectively. 

The convex conjugate functions $f^{*}$ and $g^{*}$ can be obtained
by applying the definition and rules of convex conjugate. For brevity,
we omit the details of the derivation.

\begin{align*}
f^{*}(v) & =-\sum_{i=1}^{n}v_{i}\cdot\log\left(\frac{w_{i}}{v_{i}}\right),\;v>0,\\
g^{*}(v) & =\lambda h^{*}(\frac{v}{\lambda}),\;\lambda>0,
\end{align*}
where $h^{*}$ is the convex conjugate of $||r-\hat{r}_{j}||^{p}$,
which is 
\begin{equation}
h^{*}(z)=\hat{r}_{j}^{\top}z+\frac{1}{q\cdot p^{q-1}}\cdot||z||_{*}^{q}
\end{equation}
and $q$ is the conjugate exponent to $p$, i.e., $q=\frac{p}{p-1}$.

Substituting these convex conjugate into (\ref{eq:conjugate}), we arrive at 
\begin{align*}
 & \max_{v}-f^{*}(v)-g^{*}(-v)\\
= & \max_{v}\sum_{i=1}^{n}v_{i}\log\left(\frac{w_{i}}{v_{i}}\right)-\lambda h^{*}(\frac{-v}{\lambda})\\
= & \max_{v}\sum_{i=1}^{n}v_{i}\log\left(\frac{w_{i}}{v_{i}}\right)-\lambda\left(\hat{r}_{j}^{\top}\left(\frac{-v}{\lambda}\right)+\frac{1}{q\cdot p^{q-1}}||\frac{-v}{\lambda}||_{*}^{q}\right)\\
= & \max_{v}\sum_{i=1}^{n}v_{i}\log\left(\frac{w_{i}}{v_{i}}\right)+\hat{r}_{j}^{\top}v-\frac{1}{q\cdot p^{q-1}}\lambda||\frac{v}{\lambda}||_{*}^{q}.
\end{align*}
Finally, replacing the inner minimization in (\ref{eq:dual}) by the above maximization problem and substituting $q$ with $\frac{p}{p-1}$, we obtain the final formulation.
Note that 
\begin{equation} \nonumber
\lim_{p\rightarrow1}(p-1)p^{-\left(\frac{p}{p-1}\right)}\lambda||\frac{v^{(j)}}{\lambda}||_{*}^{\frac{p}{p-1}}=\begin{cases}
0 & {\rm if}\;||v^{(j)}||_{*}\leq\lambda,\\
\infty & {\rm if}\;||v^{(j)}||_{*}>\lambda. 
\end{cases}
\end{equation}
Thus, for $p=1$, the final formulation reduces to 
\begin{equation} \nonumber
\max_{w \in {\cal W},v^{(j)}\geq0,\lambda\geq||v^{(j)}||_{*}}\;\frac{1}{N}\sum_{j=1}^{N}\left(\hat{r}_{j}^{\top}v^{(j)}+\sum_{i=1}^{n}v_{i}^{(j)}\log\left(\frac{w_{i}}{v_{i}^{(j)}}\right)\right)-\lambda\varepsilon.
\end{equation}
\end{proof}

\end{document}